\documentclass[11pt]{amsart}       
\usepackage{graphicx}

\usepackage{fullpage}

\usepackage{bm,amsfonts,color}

\newtheorem{thm}{Theorem}[section]
\newtheorem*{thm*}{Theorem}
\newtheorem{prop}[thm]{Proposition}
\newtheorem{cor}[thm]{Corollary}
\newtheorem*{cor*}{Corollary}
\newtheorem{lem}[thm]{Lemma}

\newtheorem{quest}[thm]{Question}

\theoremstyle{remark}
\newtheorem{rmk}[thm]{Remark}

\theoremstyle{remark}
\newtheorem*{rmk*}{Remark}

\theoremstyle{definition}
\newtheorem{defn}[thm]{Definition}

\theoremstyle{definition}
\newtheorem*{defn*}{Definition}

\numberwithin{equation}{section}

\newcommand{\vect}[1]{\bm{#1}}
\newcommand{\ra}{\rightarrow}

\newcommand{\sm}{\setminus}
\newcommand{\I}{1/2}
\newcommand{\R}[1]{R({#1})}
\newcommand{\M}[1]{M({#1})}
\newcommand{\area}{{\rm area}}
\newcommand{\proj}{\mathrm{proj}}
\newcommand{\genericR}{\Omega}
\newcommand{\Qconstant}{\eta}
\newcommand{\dist}{\textrm{dist}}
\newcommand{\rmin}{r_{\min}}
\newcommand{\rmax}{r_{\max}}

\begin{document}

\title{
Nearest Neighbor Distances on a Circle: Multidimensional Case
}

\author[P. ~Bleher]{Pavel M. Bleher$^1$}
\author[Y. ~Homma]{Youkow Homma$^{1,2}$}
\author[L. ~Ji]{Lyndon L. Ji$^{1,2}$}
\author[R. ~Roeder]{Roland K.\ W.\ Roeder$^1$}
\author[J. ~Shen]{Jeffrey D. Shen$^{1,3}$}

\date{\today}

\maketitle
\markboth{\textsc{P. Bleher, Y. Homma, L. Ji, R. Roeder, and J. Shen}}
  {\textit{Nearest Neighbor Distances on a Circle: Multidimensional Case}}

\footnotetext[1]{
Indiana University-Purdue University Indianapolis}
\footnotetext[2]{
Carmel High School}
\footnotetext[3]{
Park Tudor School}

\maketitle

\begin{abstract}
We study the distances, called \textit{spacings}, between pairs of neighboring
energy levels for the quantum harmonic oscillator. Specifically, we consider
all energy levels falling between $E$ and $E~+~1$, and study how the spacings
between these levels change for various choices of $E$, particularly when $E$
goes to infinity. Primarily, we study the case in which the spring constant is
a \emph{badly approximable} vector. We first give the proof by
Boshernitzan-Dyson that the number of distinct spacings has a uniform bound
independent of $E$. Then, if the spring constant has components forming a basis
of an \textit{algebraic number field}, we show that, when normalized up to a
unit, the spacings are from a finite set. Moreover, in the specific case that
the field has one fundamental unit, the probability distribution of these
spacings behaves quasiperiodically in $\log E$. We conclude by studying the
spacings in the case that the spring constant is not \emph{badly approximable},
providing examples for which the number of distinct spacings is unbounded.
\end{abstract}

\section{Introduction}

\subsection{Physical Motivation}

The quantum harmonic oscillator is given by the Hamiltonian
\begin{eqnarray}
H = -\displaystyle \sum_{j=1}^{d+1} \frac{\hbar ^2}{2m}\frac{\partial^2}{\partial x_j ^2} + \sum_{j=1}^{d+1} \frac{k_j}{2} x_j ^2. \nonumber
\end{eqnarray}

\noindent Applying Schr\"{o}dinger's equation, the quantum energy levels of the system are determined by $d+1$ non-negative integers, $m_0, \ldots, m_d$, and they are of the form
\begin{eqnarray}
E = E_0 + m_0\alpha_0 + \ldots + m_d\alpha_d, \nonumber
\end{eqnarray}
where $\alpha_0, \ldots, \alpha_d$ are positive real numbers depending on the spring constant $\vect{k}$ and the mass $m$.
The problem is to study the distribution of energy level spacings (the difference between neighboring energy levels) for all energy levels occurring in a given interval $E' \le E < E'+c$, for some fixed constant $c$, as $E' \rightarrow \infty$. It is convenient to make $c=\alpha_0$. Since
\begin{eqnarray}
E = E_0 + \alpha_0(m_0 + m_1 \omega_1 + \ldots  + m_d \omega_d), \nonumber
\end{eqnarray}
where $\omega_i = \alpha_i/\alpha_0 > 0$, the problem becomes to find spacings between energy levels
\begin{eqnarray}
\mathcal{E} = m_0 + m_1 \omega_1 + \ldots  + m_d \omega_d \nonumber
\end{eqnarray}
between a given $\mathcal{E}^\prime$ and $\mathcal{E}^\prime + 1$ as $\mathcal{E}^\prime \rightarrow \infty$. For any integer vector $\vect{m} = (m_1,\ldots, m_d)$ such that $\vect{m}\cdot \vect{\omega} < \mathcal{E}^\prime + 1$, there is exactly one $m_0 \ge 0$ that forces $\mathcal{E}$ into the interval $[\mathcal{E}',\mathcal{E}'+1)$. This allows us to reduce the problem modulo 1, considering differences between the fractional parts of the numbers $m_1 \omega_1 + \ldots + m_d\omega_d$ determined by integer vectors $\vect{m} \in R(t)$. Here, $R(t)$ is the homothetic expansion of $R = \{\vect{v} \in \mathbb{R}^n \mid v_i \geq 0, v_1\omega_1 + \ldots + v_d\omega_d < 1\}$ by a factor of $t=\mathcal{E}'+1$ about the origin.

\subsection{Mathematical Problem}\label{Mathematical Problem}

For any real number $\theta$, we will use the following notation: $\left\lfloor \theta \right\rfloor$ is the floor, $\left\lceil \theta \right\rceil$ is the ceiling, $\{\theta\} = \theta - \lfloor \theta \rfloor$ is the fractional part, and $||\theta|| = \dist(\theta, \mathbb{Z})$ is the distance between $\theta$ and the nearest integer.

Let $\omega_0=1,\omega_1,\ldots,\omega_d$, be $d+1$ real numbers, linearly independent over the rationals, $\vect{\widetilde{\omega}}=(\omega_0,\omega_1,\ldots,\omega_d)$, and $\vect{\omega} = (\omega_1,\ldots,\omega_d)$. 
Let $R$ be some bounded convex region in $\mathbb{R}^d$, $\R{t}$ be the homothetic expansion of $R$ about the origin by a factor of $t > 0$, and $\M{t}=\R{t} \cap \mathbb{Z}^d$.

We order the numbers $\{\vect{m}\cdot\vect{\omega}\}$, where $\vect{m}$ ranges over $\M{t}$, into the sequence
\begin{eqnarray}
0 \leq y_1(t) < \ldots < y_{|\M{t}|}(t) < 1. \nonumber
\end{eqnarray}
We then consider the differences between consecutive numbers in this sequence, called \emph{spacings},
\begin{eqnarray}
\delta_j(t) = y_{j}(t) - y_{j-1}(t), j=2,\ldots,|\M{t}|. \nonumber
\end{eqnarray}
 Now, let $D(t)$ be the number of distinct spacings and let $\Delta_j(t)$ be the ordered sequence of distinct spacings from the $\delta_j(t)$, so that
\begin{eqnarray}
0 < \Delta_1(t) < \ldots < \Delta_{D(t)}(t) < 1. \nonumber
\end{eqnarray}
We are interested in analyzing the set of distinct spacings as well as the frequency of occurrence of each distinct spacing from this set as $t$ increases for various choices of $\omega$ and $R$.

\begin{rmk}
The problem can also be considered on the circle $S_1=[0,1]/0~\sim~1$. However, these two formulations are equivalent up to one spacing.
\end{rmk}


\subsection{Previous Works}
In 1958, Steinhaus conjectured that in the case $d=1$, for any $w_1$ and any
$t$, there are at most three distinct spacings. This conjecture was proved in
the same year independently by P. Erd\"{o}s, G. Haj\'{o}s, V. S\'{o}s, J.
Sur\'{a}nyi, N. Swieczkowski, and P. Sz\"{u}sz.\footnote{Our source for the
history is \cite{SOS}.} Moreover, there are simple formulae for the sizes of
the three spacings at any $t$ and the frequencies with which each spacing
occurs.

In \cite{BLEHER 1}, Bleher studies the distribution of spacings as $n
\rightarrow \infty$ in the case that $d=1$ and $\omega_1 =
\frac{\sqrt{5}+1}{2}$, the golden ratio.  If one fixes $x \in (0,1]$ and
considers the distribution of spacings at the times $t_n = f_{n-1} + x
f_{n-2}$, the limit distribution exists and a concrete formula is given for the
distribution.  Meanwhile, it is shown in \cite{BLEHER 2} that for almost any
$\omega_1$ there is no limit distribution of spacings along the (analogous)
sequence of times $t_n = p_n$, where $p_n/q_n$ is $n$-th continued fraction
approximate.

Geelen and Simpson \cite{GEELEN} showed that if $d=2$ and $\omega_1$ and
$\omega_2$ are any frequencies, then using all integer points $0 \leq
v_1 < n_1$ and $0 \leq v_2 < n_2$ generates at most $n_1 + 3$ distinct
spacings. Moreover, they conjectured that for general $d$ and integer points chosen such that $0\leq v_i<n_i$, for each $i$, then the number of spacings will be no more than $\prod_{i=1}^{d-1}{n_i}+C_d$, where $C_d$ only depends on $d$.  Progress was made on this conjecture by Chevallier \cite{CHEVALLIER}, who showed that the number of spacings is no greater than $\prod_{i=1}^{d-1}{n_i}+3\prod_{i=1}^{d-2}{n_i}+1$.

Concerning the distribution of spacings in the multidimensional case, there is
a note of Dyson \cite{DYSON} and the resulting letters between Boshernitzan and
Dyson \cite{BOSHERNITZAN1,BOSHERNITZAN2}, but no other works that we are aware
of. We use the results of Boshernitzan and Dyson as a starting point of our
work, and moreover, we will present many of the details of their work, since it
was never published.

It is noteworthy that the situation is actually easier if one considers
random~${\bm \omega}$.  When $d=1$, Bleher \cite{BLEHER 2} showed that for any
random (absolutely continuous) distribution of $\omega_1$ the spacings have a
corresponding random limiting distribution that is independent of the initial
distribution of $\omega_1$.  This is a consequence of the fact that the Gauss continued fraction map
is mixing with respect to its
invariant measure.  Similar properties of the
random limit in the $d=1$ case were later studied by Greenman \cite{GREENMAN}.

Marklof \cite{MARKLOF} has proved
an exciting generalization of \cite{BLEHER 2} and \cite{GREENMAN} to
higher dimensions.  He uses the connection
between values of a linear form and the dynamics of flows on ${\rm
SL}(d,\mathbb{R})/{\rm SL}(d,\mathbb{Z})$, which has been very fruitful in the
work of Dani, Eskin, Kleinbock, Margulis, McMullen, Ratner, and others. (See,
for example, \cite{DANI}.)  It was shown by Moore  \cite{MOORE} that the flow
induced by
\begin{eqnarray*}
M \mapsto {\rm diag}\left(e^{-t},\ldots,e^{-t},e^{(d-1)t}\right)M
\end{eqnarray*}
is mixing with respect to the Haar measure on ${\rm SL}(d,\mathbb{R})/{\rm SL}(d,\mathbb{Z})$.
Marklof uses this mixing property to show that if ${\bm \omega}$ is chosen
randomly with continuous joint probability density, then each of the
$n$-point correlation densities has a limit distribution that is independent of
the initial distribution of ${\bm \omega}$.  From these $n$-point correlation densities, one can determine the
distribution of spacings.

\subsection{Main Results}\label{Main Results}
\begin{defn}\label{Badly Approximable}
A vector $\vect{\omega} \in \mathbb{R}^d$ is called \emph{Diophantine} with exponent $\gamma$ if there is some positive number $K$ such that for all nonzero vectors $\vect{m} \in \mathbb{Z}^d$, we have
\begin{eqnarray}\label{EQ:BA}
||\vect{m} \cdot \vect{\omega}|| \geq \frac{K}{|\vect{m}|^\gamma}, \textrm{ where } |\vect{m}| = \sqrt{m_1^2 + \ldots + m_d^2}.
\end{eqnarray}
It follows from the Minkowski's Theorem that $\gamma \geq d$. We call $\vect{\omega}$ \emph{badly approximable} if $\gamma = d$.
\end{defn}

The following is a result outlined in a discussion between Dyson and Boshernitzan \cite{BOSHERNITZAN2} in response to a preprint of Dyson \cite{DYSON}.

\begin{thm}\label{THM:1}\textbf{(Boshernitzan-Dyson)}
If $\vect{\omega}$ is badly approximable, then the number of distinct spacings has a bound independent of $t$.
\end{thm}

There is a very convenient way to produce badly approximable vectors $\vect{\omega}$ using algebraic number fields. If $\Phi$ is an algebraic number field and $1, \omega_1, \ldots, \omega_d$ are algebraic integers forming a basis for $\Phi$ over $\mathbb{Q}$, then a theorem of Perron \cite{PERRON} gives that $\vect{\omega} = (\omega_1,\ldots,\omega_d)$ is badly approximable. Thus, the original theorem in Dyson's preprint \cite{DYSON} is obtained as the following direct corollary to Theorem~\ref{THM:1}.

\begin{cor}\label{COR:DYSON} \textbf{(Dyson)}
If $1,\omega_1,\ldots,\omega_d$ are algebraic integers forming a basis for an algebraic number field $\Phi$, then the number of distinct spacings has a bound independent of~$t$.
\end{cor}

\noindent Since the proof of Theorem \ref{THM:1} was never published, we will present it in Section \ref{The Badly Approximable Case}, together with an explicit bound for the number of spacings. We will then focus on the case that $\vect{\omega}$ is obtained from an algebraic number field (as above), studying further questions about the spacings, including their algebraic properties, rigidity properties, and their limiting distributions.

\begin{thm}\label{THM:2}
The field norm of any spacing $\Delta_j(t)$ has a bound independent of $t$ and $j$.
\end{thm}

\begin{thm}\label{THM:3}
There exists some finite set $S=\{s_1<\ldots<s_J\} \subset \Phi$ such that every spacing $\Delta_i(t)$ has the form $us_j$ for some unit $u$ of the ring of integers $\mathbb{Z}_\Phi$ (invertible element of $\mathbb{Z}_\Phi$) and some $s_j \in S$.
\end{thm}

There is a theorem due to Dirichlet (see Theorem 38 in \cite{MARCUS}) which gives that the group of units of the ring of integers $\mathbb{Z}_\Phi$ is isomorphic to $G \times \mathbb{Z}^{r+s-1}$, where $G$ is a finite cyclic group consisting of all the roots of unity from $\Phi$, $r$ is the number of real embeddings of $\Phi$ into $\mathbb{C}$, and $s$ is the number of conjugate pairs of complex embeddings of $\Phi$ into $\mathbb{C}$. Thus, there is a set of \emph{fundamental units} $u_1,\ldots,u_l$ with $l = r+s-1$ which generates the group of units of $\mathbb{Z}_\Phi,$ modulo $G$. Note that since we are concerned only with $\vect{\omega} \in \mathbb{R}^d$, $\Phi$ is real, and $G = \{-1,1\}$.

We can say much more if $\Phi$ has only one fundamental unit, i.e. $l = 1$. By Dirichlet's Theorem, this happens if and only if $(r,s) = (2,0)$, $(1,1)$, or $(0,2)$. Again, since we deal with $\vect{\omega} \in \mathbb{R}^d$, $(r,s) = (2,0)$ or $(1,1)$.

\begin{thm}\label{THM:4''}
Suppose that the field $\Phi$ has only one fundamental unit $u$. If $l(t) = \left\lceil -d\log_u(t) \right\rceil$, then there exists a finite set $S = \{s_1< \ldots< s_J\} \subset \mathbb{Z}_\Phi$ such that $\left\{\Delta_1 (t), \ldots, \Delta_{D(t)} (t) \right\} \subseteq u^{l(t)} \cdot S$ for all $t$.
\end{thm}

\begin{cor}\label{COR:4}
If the field $\Phi$ has only one fundamental unit, then there is some finite set $\widetilde{S} \subset \Phi$ such that $\frac{\Delta_i(t)}{\Delta_1(t)} \in \widetilde{S}$ for all $i$ and $t$.
\end{cor}

\noindent We now consider the distribution of spacings in the special case that $\Phi$ has exactly one fundamental unit.

\begin{thm}\label{THM:LIMIT}
Suppose $\Phi$ is a real quadratic field having exactly one fundamental unit $0 < u < 1$ and $\vect{\omega}$ is formed by choosing a basis for $\Phi$ (over $\mathbb{Q}$) consisting of algebraic integers. Then, for any $\Qconstant \in [1, u^{-1}]$, the proportion of normalized spacings equal to $s_j$ has a limit along the subsequence $t_n=\Qconstant(\frac{1}{u})^n$, with rate of convergence equal to $\mathcal{O}(u^{2n})$.
\end{thm}

\begin{thm}\label{QUASIPERIODICITY}
Suppose $\Phi$ is a cubic field with exactly one fundamental unit $0 < u < 1$ and $\vect{\omega}$ is formed by choosing a basis for $\Phi$ (over $\mathbb{Q}$) consisting of algebraic integers. Then, for any $\Qconstant \in [1, \mu],$ where $\mu = u^{-1/2},$ there is a continuous function
\begin{eqnarray}
g : \mathbb{T} \rightarrow P = \{(p_1,\ldots,p_J) \mid \sum{p_j}=1, p_j \geq 0 \} \nonumber
\end{eqnarray}
and some frequency $\theta > 0$ such that for any $n\in\mathbb{N}$, the proportion of normalized spacings at $t_n=\Qconstant\mu^n$ equal to $s_j$ is $g_j(\theta n)+\mathcal{O}(u^{\frac{n}{2}})$.
\end{thm}

That is, we show that the distribution of normalized spacings along $t = \mu^n$ for some $\mu$ either has a limit or behaves quasiperiodically. Note that these two cases cover all number fields $\Phi$ with exactly one fundamental unit, since $(r,s) = (2,0)$ occurs only when $\Phi$ is quadratic and $(r,s) = (1,1)$ occurs only when $\Phi$ is a cubic field.

The proofs for the results above can be found in Sections~\ref{The Badly Approximable Case}-\ref{The Asymptotic Distribution of Spacings} of the paper. In the last section, we consider the case where $\vect{\omega}$ is not badly approximable. A major question is whether the converse of the Theorem~\ref{THM:1} holds.
\begin{quest}\label{QUEST:NONBADLYAPPROXIMABLE}
Is $D(t)$ unbounded if $\vect{\omega}$ is not badly approximable?
\end{quest}
\noindent Since answering this question may be difficult, we present various partial results. First, we find some examples of $\vect{\omega}$ that are not badly approximable for which $D(t)$ is unbounded by using a technique outlined in a letter from Boshernitzan to Dyson \cite{BOSHERNITZAN1}. In particular, for any $0 < \delta \leq 1$ we find vectors $\vect{\omega}$ for which $\displaystyle\limsup_{t\rightarrow\infty}\left(\frac{D(t)}{t^{1-\delta}}\right) = \infty$. Moreover, we also find an $\vect{\omega} \in \mathbb{R}^3$ that is Diophantine with exponent $5$ for which $\displaystyle\limsup_{t\rightarrow\infty}(D(t)) = \infty$. Lastly, we show that the ratio $\frac{\Delta_{D(t)}(t)}{\Delta_{1}(t)}$ between the largest and the smallest spacing is unbounded.
\vspace{0.15in}

\section{The Badly Approximable Case}\label{The Badly Approximable Case}
In this section, we will prove Theorem \ref{THM:1} using the ``Transference Theorems'' from Cassels \cite{CASSELS}.
\begin{thm}\label{THM:TRANSFERENCE}
The vector $\vect{\omega}$ is badly approximable, satisfying (\ref{EQ:BA}) with
constant $K$, if and only if for any $X_1$ and any $\alpha \in [0,1]$, there
exists a nonzero integer vector $\vect{x}$ such that $|\vect{x}|_\infty \leq
X_1$ and $|| \vect{x}\cdot\vect{\omega}-\alpha || \leq \frac{L}{X_1^d}$ ,where 
$L~=~\left(\frac{[K^{-1}]+1}{2}\right)^{d+1}K$.
\end{thm}
\begin{proof}
From (\ref{EQ:BA}), we have that for any $X > 0$ and nonzero vector $\vect{x}$ such that $|\vect{x}|_\infty < X$,
\begin{eqnarray}
|| \vect{x}\cdot\vect{\omega} || \geq \frac{K}{|x|_\infty^d} \geq \frac{K}{X^d}. \nonumber
\end{eqnarray}
 Thus, this is a direct application of  Theorem VI from Chapter V of \cite{CASSELS}, where $C = KX^{-m/n}$, $m = d$, $n = 1$, $L_1(\vect{x}) = \vect{x}\cdot\vect{\omega}$, and $X = \frac{2X_1}{\left\lfloor K^{-1}\right\rfloor+1}$.
\end{proof}

Let $A$ and $B$ be closed balls with radii $\rmin$ and $\rmax$ respectively, taken with respect to the infinity norm $|\cdot|_\infty$, such that $A \subseteq R \subseteq B$. Then let $A(t) = \{xt \mid x \in A \}$ and $B(t) = \{xt \mid x \in B\}$.


\begin{lem}\label{LEM:SMALLESTSPACING}
For all $t > 0$, the smallest spacing satisfies $\Delta_1(t) \geq \frac{K}{(2\rmax t)^d}$.
\end{lem}
\begin{proof}
Note that $\Delta_1(t) = \{\vect{n}\cdot\vect{\omega}\} -
\{\vect{m}\cdot\vect{\omega}\} = || (\vect{n}-\vect{m})\cdot\vect{\omega} ||$
for some $\vect{m},\vect{n} \in \M{t} \subset \R{t} \subset B(t)$. Then
$|\vect{n}-\vect{m}|_\infty \leq 2\rmax t$, so by (\ref{EQ:BA}), 
$|| (\vect{n}-\vect{m})\cdot\vect{\omega} || \geq \frac{K}{(2\rmax t)^d}$.
\end{proof}

If $\vect{\omega}$ is badly approximable, satisfying (\ref{EQ:BA}) with constant $K$, we set \\
$L'=(2L^{1/d}+\I)^d$, where $L = K\left(\frac{\left\lfloor K^{-1}\right\rfloor +1}{2}\right)^{d+1}$.

\begin{lem}\label{LEM:LARGESTSPACING}
For all $t > 0$, the largest spacing satisfies $\Delta_{D(t)}(t) \leq \frac{L'}{(\rmin t)^d}$.
\end{lem}
\begin{proof}
If $t$ is small, i.e. $t \leq \frac{2L^{1/d}+\I}{\rmin}$, then it is clear that $1 \leq \frac{L'}{(\rmin t)^d}$ and $\Delta_{D(t)}(t) \leq 1 \leq \frac{L'}{(\rmin t)^d}$. Otherwise if $t$ is large, i.e. $t > \frac{2L^{1/d}+\I}{\rmin}$, we will show that $\Delta_{D(t)}(t) \leq \frac{2L}{(\rmin t-\I)^d}$, which is less than $\frac{L'}{(\rmin t)^d}$ for $t > \frac{2L^{1/d}+\I}{\rmin}$.

Suppose for the sake of contradiction that $\Delta_{D(t)}(t) > \frac{2L}{(\rmin t-\I)^d}$ for some $t > \frac{2L^{1/d}+\I}{\rmin}$. Let $\Delta_{D(t)}(t) = \delta_{i}(t) = y_i(t)-y_{i-1}(t)$ for some $i$. We show that there is a $\vect{m} \in \M{t}$ such that $\{\vect{m}\cdot\vect{\omega}\}$ is in the interval $(y_{i-1}(t),y_{i}(t))$. If we let $\vect{p}$ be the integer point closest to the center of $A(t)$, let $\alpha = \frac{y_{i-1}(t)+y_{i}(t)}{2}$ be the midpoint of the interval $(y_{i-1}(t),y_{i}(t))$, and let $\beta = \alpha+\{\vect{p}\cdot\vect{\omega}\}$, then by Theorem~\ref{THM:TRANSFERENCE}, we can find an integer $\vect{x}$, and thus an integer $\vect{m} = \vect{x}+\vect{p}$, so that
\begin{eqnarray}
|\vect{m}-\vect{p}|_\infty = |\vect{x}|_\infty &\leq& \rmin t-\I \label{EQ:BALL} \\
\textrm{and } || \vect{m}\cdot\vect{\omega}-\alpha || = || \vect{x}\cdot\vect{\omega}-\beta || &\leq& \frac{L}{(\rmin t-\I)^d}. \label{EQ:INTERVAL}
\end{eqnarray}
From (\ref{EQ:BALL}), it can be seen that $\vect{m}$ is in the ball $A(t)$.
Thus, $\vect{m}$ is in $\M{t}$. Also from (\ref{EQ:INTERVAL}), the distance
from $\{\vect{m}\cdot\vect{\omega}\}$ to the midpoint $\alpha$ of the interval
$(y_{i-1}(t),y_{i}(t))$ is no more than $\frac{L}{(\rmin t-\I)^d}$, giving that
$\{\vect{m}\cdot\vect{\omega}\}$ is in the interval $(y_{i-1}(t),y_{i}(t))$.
However, this is a contradiction since $y_{i-1}(t)$ and $y_{i}(t)$ are
consecutive numbers in the ordered sequence $\{\vect{m}\cdot \vect{\omega}\},$
where $\vect{m}$ ranges over all of $\M{t}$. Therefore, we have
$\Delta_{D(t)}(t) \leq \frac{2L}{(\rmin t-\I)^d}$ when $t >
\frac{2L^{1/d}+\I}{\rmin}$, completing the proof.
\end{proof}

\begin{lem}\label{LEM:SPACINGDIFFERENCE}
For any $i$ and all $t>0$, we have $\Delta_i(t)-\Delta_{i-1}(t) \geq \frac{K}{(4\rmax t)^d}$.
\end{lem}
\begin{proof}
Note that any positive difference between spacings can be expressed as
\begin{eqnarray*}
|(\{\vect{n_1}\cdot\vect{\omega}\} -
\{\vect{m_1}\cdot\vect{\omega}\})-(\{\vect{n_2}\cdot\vect{\omega}\} -
\{\vect{m_2}\cdot\vect{\omega}\})|  
&=& || (\vect{n_1}-\vect{m_1}-\vect{n_2}+\vect{m_2})\cdot\vect{\omega} || 
\end{eqnarray*}
for some $\vect{m_1},\vect{n_1},\vect{m_2},\vect{n_2} \in \M{t} \subset B(t)$. Since 
\begin{eqnarray}
|\vect{n_1}-\vect{m_1}-\vect{n_2}+\vect{m_2}|_\infty \leq 4\rmax t, \nonumber
\end{eqnarray}
Equation (\ref{EQ:BA}) gives that $|| (\vect{n_1}-\vect{m_1}-\vect{n_2}+\vect{m_2})\cdot\vect{\omega} || \geq \frac{K}{(4\rmax t)^d}$.
\end{proof}

\noindent
{\it Proof of Theorem~\ref{THM:1}:  }
By Lemmas~\ref{LEM:SMALLESTSPACING},~\ref{LEM:LARGESTSPACING}, and \ref{LEM:SPACINGDIFFERENCE}, we have that
\begin{eqnarray}
D(t) \leq  \frac{\Delta_{D(t)}(t)-\Delta_1(t)}{\min(\Delta_i(t)-\Delta_{i-1}(t))}+1 
&\leq& \frac{\frac{L'}{(\rmin t)^d}-\frac{K}{(2\rmax t)^d}}{\frac{K}{(4\rmax t)^d}}+1 \nonumber \\  
&=& \left( \frac{4\rmax}{\rmin} \right) ^d \left( \frac{L'}{K} \right) -2^d+1. \nonumber
\end{eqnarray}
This gives an explicit bound for the number of distinct spacings.


\section{The General Algebraic Case}\label{The General Algebraic Case}
Let $1=\omega_0,\omega_1,\ldots,\omega_d$ be algebraic integers forming a basis for some number field $\Phi$.



\begin{lem}
Let $\vect{\widetilde{\omega}} = (\omega_0, \ldots, \omega_d)$ with $\omega_0 = 1$.
Perron's theorem gives a constant $K' > 0$ so that $|\vect{n} \cdot \vect{\widetilde{\omega}}| > \frac{K'}{|\vect{n}|_\infty^{d}}$. We can then find a constant $K > 0$ so that $||\vect{m}\cdot \vect{\omega}|| > \frac{K}{|\vect{m}|^d}.$
\end{lem}
\begin{proof}
First, let $\vect{n} = (n_0, m_1, \ldots, m_d)$, where $n_0$ is chosen so that $|\vect{n}\cdot\vect{\widetilde{\omega}}| = ||\vect{m}\cdot \vect{\omega}||$. It is easy to see then that $|n_0| \le |\lfloor \vect{m}\cdot\vect{\omega} \rfloor| + 1$. If $|n|_\infty = |m|_\infty \geq |m|$, then we can simply choose $ K = K'$. Otherwise, $|n|_\infty = |n_0| \le |\lfloor \vect{m}\cdot\vect{\omega} \rfloor| + 1$. Note that $|\lfloor \vect{m}\cdot\vect{\omega} \rfloor| \le |\vect{m}\cdot\vect{\omega}| + 1 \le |\vect{m}|\cdot|\vect{\omega}| + 1$, where the second half comes from the Cauchy-Schwarz inequality. From this, we have that
\begin{eqnarray}
||\vect{m}\cdot \vect{\omega}|| = |\vect{n}\cdot \vect{\widetilde{\omega}}|  > \frac{K'}{(\lfloor \vect{m}\cdot\vect{\omega} \rfloor + 1)^d} \geq \frac{K'}{(|\vect{m}||\vect{\omega}| + 2)^d} = \frac{\frac{K'}{(|\vect{\omega}| + \frac{2}{|\vect{m}|})^d}}{|\vect{m}|^d}. \nonumber
\end{eqnarray}
Since $\vect{m}$ is a non-zero integer vector, $|m| \geq 1$, so choosing $K = K'(|\vect{\omega}| + 2)^{-d}$ works.
\end{proof}

\noindent
{\it Proof of Theorem~\ref{THM:2}: }
We can write the field norm $N$ of an element $x$ from the number field $\Phi$ as
\begin{eqnarray}
N(x) = \prod_{i=0}^{d} x_j = x \prod_{i=1}^{d} x_j, \nonumber
\end{eqnarray}
where $x_j$ are the conjugate images of $x$. In other words, if 
\begin{eqnarray}
x = \vect{n} \cdot \vect{\omega} = n_0 + \ldots + n_d \omega_d, \nonumber
\end{eqnarray} 
then 
\begin{eqnarray}
x_j = n_0 + n_1 \sigma_j(\omega_1) + \ldots + n_d \sigma_j (\omega_d),\nonumber
\end{eqnarray}
where $\sigma_j: \Phi \ra \mathbb{C}$ are the conjugate embeddings. It can thus be seen that $N(x) = xQ(\vect{n})$, where $Q(\vect{n})$ is a homogeneous polynomial of degree~$d$ in the components of $\vect{n} = (n_0,\ldots,n_d)$. Then by Lemma~\ref{LEM:LARGESTSPACING}, $N(\Delta_j(t)) = \Delta_j(t)Q(\vect{n})$ is bounded above by $\frac{L'}{(at)^d}|Q(\vect{n})|$. Hence, it suffices to show that $\frac{Q(\vect{n})}{t^d}$ is bounded. This follows immediately since $Q$ is a homogeneous polynomial of degree $d$, implying the existence of a constant $C > 0$ such that $|Q(\vect{n})| \leq C |\vect{n}|_\infty ^{d} \leq C (bt)^d$.

\begin{defn}\label{Equivalence Class}
We let the \emph{equivalence class} of an element $a \in \mathbb{Z}_{\Phi}$ be
\begin{eqnarray}
E(a) = \{a u \mid u \textrm{ is a unit from } \mathbb{Z}_{\Phi}\}.
\end{eqnarray}
Note that all elements of an equivalence class have the same norm $N(au) = N(a) N(u) = N(a)$. Also note that for any $x \in \Phi$, we have that $x \in \mathbb{Z}_\Phi$ if and only if $N(x) \in \mathbb{Z}$.
\end{defn}

\begin{prop}\label{PROP:EQUIVALENCECLASSES}
For any $k \in \mathbb{Z}$, there exist elements $a_1, \ldots, a_{m(k)} \in \mathbb{Z}_\Phi$ such that $N(x) = k$ if and only if $x \in \displaystyle \bigsqcup_{i=1}^{m(k)} E(a_i)$.
\end{prop}
\begin{proof}
This follows directly from Theorem 4B of Chapter VII of \cite{SCHMIDT}.
\end{proof}

\noindent
{\it Proof of Theorem~\ref{THM:3}: }
By Theorem~\ref{THM:2}, there exists $N_0 \in \mathbb{Z}$ such that $|N(\Delta_i (t))| \leq N_0$ for all $t$ and $i$. Note that $N(\Delta_i(t))$ is an integer since $\Delta_i(t) \in \mathbb{Z}_\Phi$. Thus, by Proposition~\ref{PROP:EQUIVALENCECLASSES}, there are equivalence classes $E(a_{i,k})$, such that $\Delta_i(t) \in \displaystyle{\bigsqcup_{k=-N_0}^{N_0} \left(\bigsqcup_{i=1}^{m(k)} E(a_{i,k})\right)}$. We can thus choose the finite set
\begin{eqnarray}
S = \{a_{i,k} \mid 1 \le i \le m(k), -N_0 \le k \le N_0 \},\nonumber
\end{eqnarray}
which satisfies the conditions.


\section{The One Fundamental Unit Case}\label{The One Fundamental Unit Case}
Let the fundamental unit $u$ be chosen so that $0 < u < 1$. Note that every unit of $\mathbb{Z}_\Phi$ is of the form $\pm u^n$, where $n\in \mathbb{Z}$.

We first present a lemma necessary to the proof of Theorem \ref{THM:4''}.

\begin{lem}\label{LEM:NORMSET}
Let $k$ be any integer and $0 < x_0 < x_1 < \infty$. Then the set given by
\begin{eqnarray}\label{EQ:NORMSET}
\left\{x\in\mathbb{Z}_{\Phi} \mid N(x) = k\right\} \cap \left[x_0, x_1\right]
\end{eqnarray}
is finite.
\end{lem}
\begin{proof}

First, if $E(a_i) = \left\{a_i u^{n}\right\}$ denotes the equivalence class generated by $a_i \in \mathbb{Z}_{\Phi},$ then the set $E(a_i)\cap \left[x_0, x_1\right]$ is finite. To see this, note that if \\ $x_0 \leq a_i u^n \leq x_1$, then
\begin{eqnarray}
        \frac{\log x_0 - \log a_i}{\log u} \leq n \leq \frac{\log x_1 - \log a_i}{\log u}. \nonumber
\end{eqnarray}
 Now, by Proposition \ref{PROP:EQUIVALENCECLASSES}, we have that
\begin{eqnarray}
\left\{x\in\mathbb{Z}_{\Phi} \mid N(x) = k\right\} \cap \left[x_0, x_1\right] = \bigsqcup_{i=1}^{m(k)} E(a_i) \cap \left[x_0,x_1\right]. \nonumber
\end{eqnarray}
Since this is equal to $\displaystyle\bigsqcup_{i=1}^{m(k)} (E(a_i) \cap \left[x_0,x_1\right])$, (\ref{EQ:NORMSET}) is finite.
\end{proof}

\noindent
{\it Proof of Theorem \ref{THM:4''}: }
By Theorem~\ref{THM:2}, there is a $N_0$ such that the norm $|N(\Delta_j(t))| \leq N_0$ for all $j$ and $t$. Recall that $N(\Delta_j(t))$ is always an integer. Furthermore, by Lemmas~\ref{LEM:SMALLESTSPACING}~and~\ref{LEM:LARGESTSPACING} and by the definition of $l(t)$,
\begin{eqnarray}
x_0 u^{l(t)} \leq \frac{K}{(2bt)^d} \leq \Delta_j(t) \leq \frac{L'}{(at)^d} \leq x_1 u^{l(t)} \nonumber
\end{eqnarray}
where $x_0 = \frac{K}{(2b)^d}$ and $x_1 = \frac{L'}{a^d}u$. Therefore, we have that
\begin{eqnarray}
\Delta_j(t) \in S(t) = \bigsqcup_{i=-N_0}^{N_0}\left\{x \in \mathbb{Z}_\Phi \mid N(x) = i \right\} \cap [x_0 u^{l(t)},x_1 u^{l(t)}] \nonumber
\end{eqnarray}
for all $j$ and $t$. Note that $N(x) = N(u^{-l(t)}x)$. Thus, each $\Delta_j (t) \in u^{l(t)} S$ for the set
\begin{eqnarray}
S = \bigsqcup_{i=-N_0}^{N_0}\left\{x \in \mathbb{Z}_\Phi \mid N(x) = i \right\} \cap [x_0,x_1], \nonumber
\end{eqnarray}
which is finite by Lemma~\ref{LEM:NORMSET}.

\section{The Asymptotic Distribution of Spacings}\label{The Asymptotic Distribution of Spacings}

Let $\Phi$ be a number field with exactly one fundamental unit $0 < u < 1$. In
this section, we will prove
Theorems~\ref{THM:LIMIT}~and~\ref{QUASIPERIODICITY}. We will first prove
Theorem~\ref{QUASIPERIODICITY}, because Theorem~\ref{THM:LIMIT} follows from
some simple modifications of its proof. Thus, we first assume that $\Phi$ is a
cubic field, in which case $\Phi$ has one real embed ding and two complex
embeddings.  We first define two vector representations of a number in $\Phi$.

\begin{defn}\label{Vector Representation}
Let $\vect{n}: \Phi \ra \mathbb{Q}^{3}$ be the vector expansion an element of $\Phi$ in terms of the basis $\vect{\widetilde{\omega}} = (1,\omega_1,\omega_2)$ and $\vect{m}: \Phi \ra \mathbb{Q}^
{2}$ be such that if $\vect{n}(y) = (m_0,m_1,m_2)$, then $\vect{m}(y) = (m_1,m_2)$.
\end{defn}
\subsection{Fundamental Circle}
Let $u_1$ and $u_2 = \overline{u_1}$ be the Galois conjugates of $u$.

\begin{lem}\label{LEM:PROJDIST}
There is a two-dimensional linear subspace $E^u\in \mathbb{R}^3$ containing a circle $\mathbb{T}$ of radius $r \equiv r(\Qconstant)$ and a constant $C > 0$ such that  $\dist\left(\frac{\vect{n}(u
^n)}{t_n}, \mathbb{T}\right)~\leq~Cu^n$.
\end{lem}

\begin{proof}
Since $u$ is an irrational unit in the cubic field $\Phi$, it is a zero of an irreducible monic integral polynomial
\begin{eqnarray}
P(x) = x^3+a_2x^2+a_1x+a_0.
\end{eqnarray}
Hence,
\begin{eqnarray}
u^3 = -a_2u^2-a_1u-a_0. \nonumber
\end{eqnarray}

Let us choose our basis for $\Phi$ to be $(1,u,u^2)$. Then in this basis, multiplication by $u$ is represented by the matrix
\begin{eqnarray}
U= \left(\begin{array}{ccc}0 & 0 & -a_0 \\1 & 0 & -a_1 \\
0 & 1 & -a_2 \end{array} \right). \nonumber
\end{eqnarray}
The characteristic polynomial of $U$ is precisely the minimal polynomial of
$u$, $P(x)$. Thus, $u$, $u_{1}$, and $u_2$ are the eigenvalues of $U$. Let
$E^s$ and $E^u$ be the stable and unstable eigenspaces of $U$, respectively. We
know that $N(u)=1=a_0=uu_{1}u_{2}$ and $0<u<1$, so $u_{1}u_{2}>1$. Because
$u_{1}$ and $u_{2}$ are complex conjugates,
$\left|u_{1}\right|=\left|u_{2}\right| $. Hence, $\left|u_{1}\right|$,
$\left|u_{2}\right| > 1$.

Set $\vect{n}(1)= \vect{v_{s}}+\vect{v_{u}},$ where $\vect{v_{s}}$ is in $E^s$ corresponding to the eigenvalue $u$ and $\vect{v_{u}}$ is in $E^u$ corresponding to the eigenvalues $u_{1}$ and $u_
{2}$. Then the norms of the components of $\vect{n}(u^{n})$ in those eigenspaces are $|u|^{n}|\vect{v_{s}}|$ and $|u_{1}|^{n}|\vect{v_{u}}|,$ respectively. A calculation shows that $\vect{n}(1)$
 is not an eigenvector of $U$ and thus, $v_u$ is nonzero. Consequently, $|\frac{\vect{n}(u^{n})}{t_n}|$ does not go to zero.

Using our previous definitions, we get
\begin{eqnarray}  \left\|\frac{\vect{n}(u^n)}{t_n}\right\|=\left\|\frac{U^n\vect{n}(1)}{t_n}\right\|=\left\|\frac{U^n(\vect{v_u}+\vect{v_s})}{t_n}\right\|. \nonumber
\end{eqnarray}

\noindent Since $u u_1 u_2 = 1$ and $|u_1| = |u_2|$, we must have that $|u_1| = \mu$. Therefore,
\begin{eqnarray}
\left\|\frac{U^n \vect{v_u}}{t_n}\right\|=\frac{\left|u_{1}^n\right|\cdot\left\|\vect{v_u}\right\|}{\mu^n}= \left\|\vect{v_u}\right\| =: r. \nonumber
\end{eqnarray}

\noindent Moreover,
\begin{eqnarray}
        \left\|\frac{U^n(\vect{v_u}+\vect{v_s})}{t_n}-\frac{U^n \vect{v_u}}{t_n}\right\|=\left\|\frac{U^n \vect{v_s}}{t_n}\right\|\leq\left\|U^n \vect{v_s}\right\|\leq Cu^n. \nonumber
\end{eqnarray}

\noindent Thus, $\frac{\vect{n}(u^n)}{t_n}$ converges to $\mathbb{T}$ with rate $\mathcal{O}(u^n)$. \end{proof}

\subsection{Partitions}
\begin{figure}[hbtp]
\begin{center}


\begin{picture}(0,0)%
\includegraphics{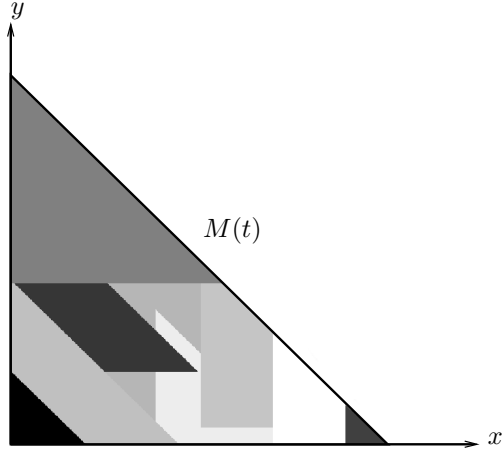}%
\end{picture}%
\setlength{\unitlength}{3947sp}%
\begingroup\makeatletter\ifx\SetFigFont\undefined%
\gdef\SetFigFont#1#2#3#4#5{%
  \reset@font\fontsize{#1}{#2pt}%
  \fontfamily{#3}\fontseries{#4}\fontshape{#5}%
  \selectfont}%
\fi\endgroup%
\begin{picture}(3052,2895)(4148,-4043)
\put(5398,-2683){\makebox(0,0)[lb]{\smash{{\SetFigFont{10}{12.0}{\familydefault}{\mddefault}{\updefault}{\color[rgb]{0,0,0}$M(t)$}%
}}}}
\put(7185,-3988){\makebox(0,0)[lb]{\smash{{\SetFigFont{10}{12.0}{\familydefault}{\mddefault}{\updefault}{\color[rgb]{0,0,0}$x$}%
}}}}
\put(4190,-1283){\makebox(0,0)[lb]{\smash{{\SetFigFont{10}{12.0}{\familydefault}{\mddefault}{\updefault}{\color[rgb]{0,0,0}$y$}%
}}}}
\end{picture}%

\caption{The partition $\{M_k(t)\}$ of $M(t)$ for $\vect{\omega} = (\sqrt[3]{2}, \sqrt[3]{4})$ and $t = 650$.}
\label{FIG:PARTITION}
\end{center}
\end{figure}

Let $Y_k(t)$ be the set of numbers $y_j(t)$ such that $\delta_j(t) = y_j(t) - y_{j-1}(t) = s_ku^{l(t)}$ and $M_k(t)$ to be the set of vectors $\vect{m} \in M(t)$ such that $\{\vect{m}\cdot\vect{\omega}\} \in Y_k(t)$. Note that $\displaystyle\bigsqcup_{k=1}^{J}{M_k(t)}= M(t)$ up to the point corresponding to $y_1(t)$. Thus, we call $\{M_k(t)\}$ a ``partition'' of $M(t)$ (see Figure~\ref{FIG:PARTITION}). Also note that $|M_k(t)|$ is the number of spacings that are equal to $s_ku^{l(t)}$. Therefore, $\frac{|M_k(t)|}{|M(t)|-1}$ is the proportion of spacings that is equal to $s_ku^{l(t)}$.

\begin{prop}{} \label{PROP:PARTITIONFORMULA}
Let $\vect{v_k} = \vect{m}(s_ku^{l(t)})$. Then, for $k=1,\ldots,J$, we have
\begin{eqnarray}
M_k(t) = [M(t)\cap(M(t)+\vect{v_k})]\sm\bigcup_{i=1}^{k-1}(M(t)+\vect{v_i}) \label{EQ:PARTITIONTWO},
\end{eqnarray}
\noindent
where $\genericR+\vect{v}$ is defined to be $\{\vect{u}+\vect{v} \mid \vect{u} \in \genericR\}$ for any $\genericR \subseteq \mathbb{R}^d$ and $\vect{v} \in \mathbb{R}^d$.
\end{prop}
\begin{proof}
We first prove that
\begin{eqnarray}
M_k(t) = [M(t)\cap(M(t)+\vect{v_k})]\sm\bigsqcup_{i=1}^{k-1}{M_i(t)} \label{EQ:PARTITIONONE}.
\end{eqnarray}
Suppose we are given some $\vect{m} \in [M(t)\cap(M(t)+\vect{v_k})]\sm\bigsqcup_{i=1}^{k-1}{M_i(t)}$. Since $\vect{m} \in M(t)$, there exists $n$ such that $y_{n}(t) = \{\vect{m}\cdot\vect{\omega}\}$. We will show that $\delta_n(t) = s_ku^{l(t)}$. For every $j < k$, $\vect{m} \notin M_j(t)$ and thus $\delta_{n}(t) \neq s_ju^{l(t)}$. If $\delta_n(t) = s_ju^{l(t)}$ with $j > k$, a contradiction is reached since
\begin{eqnarray}
\delta_n(t) = s_ju^{l(t)} > s_ku^{l(t)} = y_n(t)-\{(\vect{m}-\vect{v_k})\cdot\vect{\omega}\}, \nonumber
\end{eqnarray}
where $(\vect{m}-\vect{v_k}) \in M(t)$ because $\vect{m} \in (M(t)+\vect{v_k})$. Thus, $\delta_{n}(t) = s_ku^{l(t)}$ and therefore $\vect{m} \in M_k(t)$.

Suppose instead that $\vect{m} \in M_k(t)$. Clearly $\vect{m} \notin M_j(t)$
for $j < k$. Now let $n$ be such that $y_n(t) =
\{\vect{m}\cdot\vect{\omega}\}$. Then $y_{n-1}(t) = y_{n}(t)-s_ku^{l(t)} =
\{(\vect{m}-\vect{v_k})\cdot\vect{\omega}\}$ and thus $(\vect{m}-\vect{v_k})
\in M(t)$. Therefore, $\vect{m} \in M(t)+\vect{v_k}$ and since $\vect{m} \in
M(t)$ and $\vect{m} \notin \bigsqcup_{i=1}^{k-1}{M_i(t)}$, we have that
\begin{eqnarray*}
\vect{m}\in[M(t)\cap(M(t)+v_k)]\sm\bigsqcup_{i=1}^{k-1}{M_i(t)}.
\end{eqnarray*}
Therefore \ref{EQ:PARTITIONONE} holds. Using induction, it can be shown that
\begin{eqnarray}
\bigsqcup_{i=1}^{k-1}{M_i(t)} = \bigcup_{i=1}^{k-1}(M(t)+\vect{v_i})\cap{M(t)}, \nonumber
\end{eqnarray}
which proves (\ref{EQ:PARTITIONTWO}).
\end{proof}

Denote the power set of $R$ by $\mathcal{P}(R)$. Then let $P~:~(\mathbb{R}^2)^J~\ra~\mathcal{P}(R)^J$ map a vector of vectors $\vect{v} = (\vect{v_1},\ldots,\vect{v_J})$ to $(P_1(\vect{v}),\ldots,P_J(\vect{v}))$, where
\begin{eqnarray}\label{EQ:PARTITIONP}
P_k(\vect{v}) = [R\cap(R+\vect{v_k})]\sm\bigcup_{i=1}^{k-1}(R+\vect{v_i}).
\end{eqnarray}

\begin{prop}\label{PROP:APPROXP}
If $\vect{v} = \left(\frac{\vect{m}(s_1u^{l(t)})}{t},\ldots,\frac{\vect{m}(s_Ju^{l(t)})}{t}\right)$, then
\begin{eqnarray}\label{EQ:APPROXP}
\frac{\area(P_k(\vect{v}))}{\area(R)} - \frac{|M_k(t)|}{|M(t)|-1} = \mathcal{O}\left(\frac{1}{t}\right).
\end{eqnarray}
\end{prop}
\begin{proof}
If we let $\vect{x} = t\vect{v} = (\vect{m}(s_1u^{l(t)}),\ldots,\vect{m}(s_Ju^{l(t)}))$, then Proposition \ref{PROP:PARTITIONFORMULA}, followed by the inclusion-exclusion principle, gives that
\begin{eqnarray}
|M_k(t)| &=& \left|[M(t)\cap(M(t)+\vect{x_k})]\sm\bigcup_{i=1}^{k-1}(M(t)+\vect{x_i})\right|  \label{EQ:MKTPIE} \\
&=& \left|M(t)\cap(M(t)+\vect{x_k})\right|-\left|\bigcup_{i=1}^{k-1}[(M(t)+\vect{x_i})\cap{M(t)}\cap(M(t)+\vect{x_k})]\right| \nonumber \\
&=& \left|M(t)\cap(M(t)+\vect{x_k})\right| \nonumber \\
&\qquad&-\sum_{I \subseteq \{1,\ldots,k-1\}}(-1)^{|I|}\left|\bigcap_{i \in I}(M(t)+\vect{x_i})\cap{M(t)}\cap(M(t)+\vect{x_k})\right|.  \nonumber
\end{eqnarray}
Similarly,
\begin{eqnarray}
\area(tP_k(\vect{v})) &=& \area(tR\cap(tR+t\vect{v_k})) \label{EQ:PKVPIE} \\
&-&\sum_{I \subseteq \{1,\ldots,k-1\}}(-1)^{|I|}\area\left(\bigcap_{i \in I}(tR+t\vect{v_i})\cap{tR}\cap(tR+t\vect{v_k})\right). \nonumber
\end{eqnarray}
Given any bounded convex region $\genericR \in \mathbb{R}^2$ and the corresponding set $M = t\genericR \cap \mathbb{Z}^2$, it is well-known that $\area(t\genericR)-|M|= \mathcal{O}(t)$. Thus, by summing over all the parts of (\ref{EQ:PKVPIE}), and subtracting all the corresponding parts in (\ref{EQ:MKTPIE}), we have that
\begin{eqnarray}
\area(tP_k(\vect{v}))-|M_k(t)| = \mathcal{O}(t). \label{EQ:APPROXP1}
\end{eqnarray}
Also, $|M(t)| - \area(R(t)) = \mathcal{O}(t)$ implies
\begin{eqnarray}
 \frac{|M_k(t)|}{\area(R(t))}&-&\frac{|M_k(t)|}{|M(t)|-1} = 
 \frac{|M_k(t)|}{|M(t)|-1}\left(\frac{|M(t)|-1-\area(R(t))}{\area(R(t))}\right) = \mathcal{O}\left(\frac{1}{t}\right). \label{EQ:APPROXP2}
\end{eqnarray}
Dividing (\ref{EQ:APPROXP1}) by $\area(R(t))$ and adding (\ref{EQ:APPROXP2}) proves the statement.
\end{proof}
\begin{prop}\label{PROP:LIPSCHITZP}
The function $P$ is Lipschitz continuous with respect to the infinity norm on $(\mathbb{R}^2)^J$ and the metric $d(P^{(1)},P^{(2)}) = \sum_{i=1}^{J}{\area(P^{(1)}_i \triangle P^{(2)}_i)}$, where $\triangle$ denotes the symmetric difference between two sets.
\end{prop}
In order to prove Proposition~\ref{PROP:LIPSCHITZP}, we first present a lemma.
\begin{lem}\label{LEM:LIPSCHITZP}
Let $\genericR$ be a bounded convex region in $\mathbb{R}^2$ with a nonempty interior. Then the function $F$ taking every vector $\vect{v} \in \mathbb{R}^2$ to the set of points $\genericR+\vect{v}$ is Lipschitz.
\end{lem}
\begin{proof}
Suppose we have two vectors $\vect{v_1}$, $\vect{v_2}$ such that $|\vect{v_1}-\vect{v_2}|_\infty = x$. If $x > 1$, then
\begin{eqnarray}
\area((\genericR+\vect{v_1})\triangle(\genericR+\vect{v_2})) \leq 2[\area(\genericR)] < 2[\area(\genericR)]x. \nonumber
\end{eqnarray}
Otherwise, assume that $x \leq 1$. If $a$ is the radius, taken with respect to the infinity norm, of a closed ball contained inside $\genericR+\vect{v_1}$, then we construct $\genericR^{12}$ to be the homothetic expansion of $\genericR+\vect{v_1}$ about the center of the ball by $1+\frac{x}{a}$. It can be shown that $\genericR^{12}$ contains all vectors within $x$ of $\genericR+\vect{v_1}$.
Thus, $(\genericR+\vect{v_1})\cup (\genericR+\vect{v_2}) \subseteq \genericR^{12}$. Therefore,
\begin{eqnarray}
\area((\genericR+\vect{v_1})\triangle(\genericR+\vect{v_2})) &=& 2[\area((\genericR+\vect{v_1})\cup(\genericR+\vect{v_2}))-\area(\genericR)]  \nonumber \\
&\leq& 2[\area(\genericR^{12})-\area(\genericR)] \nonumber \\
&=& 2\left[\area(\genericR)\left(\left(1+\frac{x}{a}\right)^2-1\right)\right]. \nonumber
\end{eqnarray}
The result follows since $Q(x) = (1+\frac{x}{a})^2-1$ is Lipschitz on the interval $[0,1]$.
\end{proof}

\noindent
{\it Proof of Proposition~\ref{PROP:LIPSCHITZP}: }
From Lemma~\ref{LEM:LIPSCHITZP}, the function $F_1$ taking every vector $\vect{v} \in \mathbb{R}^2$ to the set of points $R$ is Lipschitz. It is easy to see that $F_2$ taking $\vect{v} \in \mathbb{R}^2$ to $R\cap(R+\vect{v})$ is also Lipschitz. It can be readily proven that

\begin{eqnarray}
(A\sm C) \triangle (B \sm D) &\subseteq (A \triangle B) \cup (C \triangle D), \label{EQ:SETTHEOREMONE}\\
(A \cup B) \triangle (C \cup D) &\subseteq (A \triangle C) \cup (B \triangle D) \label{EQ:SETTHEOREMTWO}
\end{eqnarray}
for any sets $A,B,C,D$. Substituting $F_1$ and $F_2$ into the formula for $P_i(\vect{v})$ in (\ref{EQ:PARTITIONP}), we have that
{\footnotesize
\begin{eqnarray}
\area(P_i(\vect{v^{(1)}}) \triangle P_i(\vect{v^{(2)}})) 
&=& \area\left(\left[F_2(\vect{v^{(1)}_i})\sm\bigcup_{j=1}^{i-1}F_1(\vect{v^{(1)}_j})\right]\triangle\left[F_2(\vect{v^{(2)}_i})\sm\bigcup_{j=1}^{i-1}F_1(\vect{v^{(2)}_j})\right]\right)  \nonumber \\
&\stackrel{(\ref{EQ:SETTHEOREMONE})}{\leq}& \area(F_2(\vect{v^{(1)}_i})\triangle F_2(\vect{v^{(2)}_i})) + \area\left(\bigcup_{j=1}^{i-1}F_1(\vect{v^{(1)}_j}) \triangle \bigcup_{j=1}^{i-1}F_1(\vect{v^{(2)}_j})\right) \nonumber \\
&\stackrel{(\ref{EQ:SETTHEOREMTWO})}{\leq}& \area(F_2(\vect{v^{(1)}_i})\triangle F_2(\vect{v^{(2)}_i})) + \sum_{j=1}^{i-1}\area(F_1(\vect{v^{(1)}_j})\triangle F_1(\vect{v^{(2)}_j})) \nonumber \\
&\leq& (C_1+(i-1)C_2)|\vect{v^{(1)}}-\vect{v^{(2)}}|_\infty \nonumber
\end{eqnarray}
}
since $F_1$ and $F_2$ are Lipschitz. Therefore, $P$ is also Lipschitz.

\subsection{Proof of Theorem~\ref{QUASIPERIODICITY} }
\begin{proof}
Let $\mathbb{T}$ be the circle with radius $r = |\vect{v_u}|$ centered at the origin on the unstable eigenspace, $E^{u}$, let $L = \{\mathrm{span}(\omega_1,\omega_2)\}$, and let $g: \mathbb{R}^3 \mapsto \mathbb{R}^{J}$ such that $g = h\circ P$, where
\begin{eqnarray}
&h: \vect{w} \longmapsto \langle \proj_{L}(S_1 \vect{w}), \ldots, \proj_{L}(S_J \vect{w})\rangle \nonumber \\
&\hspace{1.2in}\textrm{and} \hspace{1.2in} \nonumber \\
&P: \vect{v} \longmapsto (P_1(\vect{v}), \ldots, P_J(\vect{v})). \nonumber
\end{eqnarray}
Since $h$ is a composition of two Lipschitz continuous functions, projection and matrix multiplication, it is also Lipschitz continuous. Combined with Lemma \ref{LEM:PROJDIST}, this gives
\begin{eqnarray}
\vline \hspace{.05in} \vline \hspace{.02in} h\left(\proj_{E^u} \left(\frac{\vect{n}(u^{n})}{t_n}\right)\right) - h\left(\frac{\vect{n}(u^{n})}{t_n}\right) \vline \hspace{.05in} \vline < C\upsilon^{n}, \nonumber
\end{eqnarray}
where $0 < \upsilon = \frac{u}{|u_1|} = u^{3/2} < u^{1/2} < 1$. By Proposition \ref{PROP:LIPSCHITZP}, $P$ is also Lipschitz, so
\begin{eqnarray}\label{EQ:AREADIFF}
d\left(P\circ h\left(\proj_{E^u} \frac{\vect{n}(u^{n})}{t_n}\right),P\circ h\left(\frac{\vect{n}(u^{n})}{t_n}\right)\right) < C'\upsilon^{n}
\end{eqnarray}
for some constant $C'$.

Let $\vect{w_1}, \vect{w_2}$ be an orthonormal basis on $E^u$ such that $\proj_{E^u}(\vect{n}(1)) = \proj_{E^u} (1,0,0)$ coincides with $\vect{w_1}$.
The vector $\proj_{E^u} \left(\frac{\vect{n}(u^{n})}{t_n}\right)$ will thus
form an angle $\theta n$ with $\vect{w_1}$, where $\theta = \arg(u_1)$. By
Proposition \ref{PROP:APPROXP}, the area of the $j^{th}$ component of $P\circ
h\left(\frac{\vect{n}(u^{n})}{t_n}\right)$ differs from the true proportion by
at most $\frac{K}{\mu^{n}} = Ku^{n/2}$ for a constant $K$. Hence, the total
error is given by $Ku^{n/2}$ from Proposition \ref{PROP:APPROXP} combined with
$C'\upsilon^{n}$ from (\ref{EQ:AREADIFF}). Therefore, the proportion of
normalized spacings equal to $s_i$ is given by $g_i (\theta n) +
\mathcal{O}(\mu^{-n}) = g_i (\theta n)~+~\mathcal{O}(u^{n/2})$.
\end{proof}

\subsection{Proof of Theorem~\ref{THM:LIMIT} }
\begin{proof}
Using the same arguments as in Lemma~\ref{LEM:PROJDIST} gives that $u$ and $\frac{1}{u}$ are the eigenvalues of $U$. A quick calculation shows that $\frac{\vect{n}(U^n(\vect{\upsilon_u}))}{t_n} = \frac{\vect{\upsilon_u}}{\eta}$ and $\frac{\vect{n}(U^n(\vect{\upsilon_s}))}{t_n} = u^{2n}\frac{\vect{\upsilon_v}}{\eta}$. Thus, $\frac{\vect{n}(u^n)}{t_n}$ converges to the point $\vect{r} = \frac{\vect{\upsilon_u}}{\eta}$ at a rate of $\mathcal{O}(u^{2n})$. Furthermore, Proposition~\ref{PROP:LIPSCHITZP} still holds for $d=1$ so long as there is one fundamental unit. Therefore, the same techniques in the proof of Theorem~\ref{QUASIPERIODICITY} give the desired result.
\end{proof}


\section{The Non-Badly Approximable Case}
We first provide examples of non-badly approximable vectors for which the number of distinct spacings $D(t)$ goes to infinity, following
a construction that Boshernitzan outlined in a letter to Dyson, \cite{BOSHERNITZAN2}. The following Lemma will be central in both constructions:

\begin{figure}[hbtp]
\begin{center}
\scalebox{0.75}{ 


\begin{picture}(0,0)%
\includegraphics{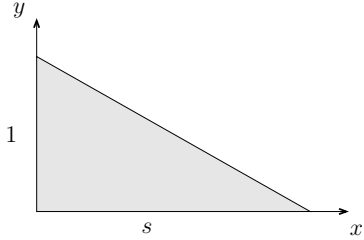}%
\end{picture}%
\setlength{\unitlength}{3947sp}%
\begingroup\makeatletter\ifx\SetFigFont\undefined%
\gdef\SetFigFont#1#2#3#4#5{%
  \reset@font\fontsize{#1}{#2pt}%
  \fontfamily{#3}\fontseries{#4}\fontshape{#5}%
  \selectfont}%
\fi\endgroup%
\begin{picture}(2915,2107)(1556,-2497)
\put(1637,-573){\makebox(0,0)[lb]{\smash{{\SetFigFont{12}{14.4}{\familydefault}{\mddefault}{\updefault}{\color[rgb]{0,0,0}$y$}%
}}}}
\put(1571,-1650){\makebox(0,0)[lb]{\smash{{\SetFigFont{12}{14.4}{\familydefault}{\mddefault}{\updefault}{\color[rgb]{0,0,0}$1$}%
}}}}
\put(2711,-2415){\makebox(0,0)[lb]{\smash{{\SetFigFont{12}{14.4}{\familydefault}{\mddefault}{\updefault}{\color[rgb]{0,0,0}$s$}%
}}}}
\put(4456,-2428){\makebox(0,0)[lb]{\smash{{\SetFigFont{12}{14.4}{\familydefault}{\mddefault}{\updefault}{\color[rgb]{0,0,0}$x$}%
}}}}
\end{picture}%

}
\caption{A closed triangle bounded by lines $x=0$, $y=0$ and $x+sy=s$.}
\label{FIG:RTRIANGLE}
\end{center}
\end{figure}

\begin{lem}\label{LEM:INFSPACING}
Let $R$ be the triangle shown in Figure~\ref{FIG:RTRIANGLE} for some $s > 0$.
Let $\frac{p}{q} \in \mathbb{Q}$ satisfy $\frac{p}{q} \leq s$, $\alpha$ be
Diophantine with exponent $\gamma$, satisfying~(\ref{Badly Approximable}) with
constant $K$, and $\vect{\omega} = \left(\alpha,
\alpha(\frac{p}{q}+\epsilon)\right)$ for some $\epsilon > 0$. Then, for all 
\begin{eqnarray}
t < t_0 = \left(\frac{K}{q^{1+\gamma}(2s)^{\gamma}\alpha\epsilon}\right)^{\frac{1}{1+\gamma}}, \nonumber
\end{eqnarray}
the number of distinct spacings satisfies $D(t) \geq \lfloor t/q \rfloor$. Here
we consider points on the circle $S_1 = [0,1]/0\sim1$.
\end{lem}
\begin{proof}
Note that all points of the form $\left\{(m,n) \cdot \vect{\omega}\right\}$ around the circle can be expressed as the sum\footnote{Sometimes, the sum of the two components is at least $1$, so the sum may sometimes need to be taken modulo $1$.} of two ``components,'' $\left\{m(\alpha)+n(\frac{p}{q}\alpha)\right\}$ and $\left\{n\alpha\epsilon\right\}$. Consider all distinct points of the form $\left\{m(\alpha)+n(\frac{p}{q}\alpha)\right\} = \left\{\frac{l}{q}\alpha\right\}$, with $(m, n) \in M(t)$. Then, let the distinct spacings for this set of points be ordered $\Delta_1(t) < \Delta_2(t) < \ldots < \Delta_N(t)$. Note that $\alpha/q$ is Diophantine with exponent $\gamma$ and constant $\frac{K}{q}$ since $||m(\alpha/q)|| \geq \frac{1}{q}||m\alpha||$. Thus, since $0 \leq l \leq stq$, we have $\Delta_j(t) = ||(l_1-l_2)\frac{\alpha}{q}|| \geq \frac{K}{(stq)^{\gamma}q} > n\epsilon\alpha$. Note also that an integer point $(m,n) \in M(t)$ has first component $\frac{l}{q}\alpha$ if and only if it is on the line $mq+np=l$ (see Figure~\ref{FIG:RTRIANGLELINES}). Thus, the points $\left\{(m,n) \cdot \vect{\omega}\right\}$ with the same first component lie consecutively on the circle as $n$ increases (see Figure~\ref{FIG:POINTSCIRCLE}).

\begin{figure}[hbtp]
\begin{center}
\scalebox{0.8}{


\begin{picture}(0,0)%
\includegraphics{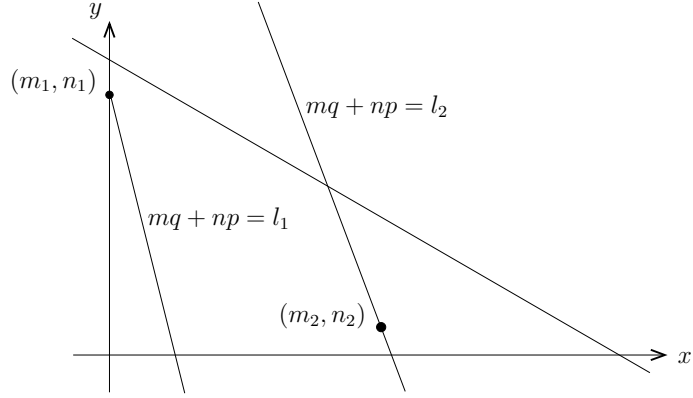}%
\end{picture}%
\setlength{\unitlength}{3947sp}%
\begingroup\makeatletter\ifx\SetFigFont\undefined%
\gdef\SetFigFont#1#2#3#4#5{%
  \reset@font\fontsize{#1}{#2pt}%
  \fontfamily{#3}\fontseries{#4}\fontshape{#5}%
  \selectfont}%
\fi\endgroup%
\begin{picture}(5275,3170)(884,-4264)
\put(3011,-3692){\makebox(0,0)[lb]{\smash{{\SetFigFont{12}{14.4}{\familydefault}{\mddefault}{\updefault}{\color[rgb]{0,0,0}$(m_2,n_2)$}%
}}}}
\put(6144,-4018){\makebox(0,0)[lb]{\smash{{\SetFigFont{12}{14.4}{\familydefault}{\mddefault}{\updefault}{\color[rgb]{0,0,0}$x$}%
}}}}
\put(1522,-1273){\makebox(0,0)[lb]{\smash{{\SetFigFont{12}{14.4}{\familydefault}{\mddefault}{\updefault}{\color[rgb]{0,0,0}$y$}%
}}}}
\put(1981,-2930){\makebox(0,0)[lb]{\smash{{\SetFigFont{12}{14.4}{\familydefault}{\mddefault}{\updefault}{\color[rgb]{0,0,0}$mq+np=l_1$}%
}}}}
\put(3223,-2055){\makebox(0,0)[lb]{\smash{{\SetFigFont{12}{14.4}{\familydefault}{\mddefault}{\updefault}{\color[rgb]{0,0,0}$mq+np=l_2$}%
}}}}
\put(899,-1850){\makebox(0,0)[lb]{\smash{{\SetFigFont{12}{14.4}{\familydefault}{\mddefault}{\updefault}{\color[rgb]{0,0,0}$(m_1,n_1)$}%
}}}}
\end{picture}%

}

\caption{The points $(m_1,n_1)$ and $(m_2,n_2)$ on lines $mq+np=l_1$ and $mq+np=l_2$.}
\label{FIG:RTRIANGLELINES}
\end{center}
\end{figure}

\begin{figure}[hbtp]
\begin{center}
\scalebox{0.75}{

\begin{picture}(0,0)%
\includegraphics{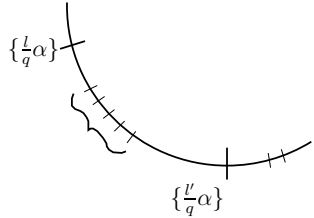}%
\end{picture}%
\setlength{\unitlength}{3947sp}%
\begingroup\makeatletter\ifx\SetFigFont\undefined%
\gdef\SetFigFont#1#2#3#4#5{%
  \reset@font\fontsize{#1}{#2pt}%
  \fontfamily{#3}\fontseries{#4}\fontshape{#5}%
  \selectfont}%
\fi\endgroup%
\begin{picture}(2573,1786)(3031,-3834)
\put(4407,-3756){\makebox(0,0)[lb]{\smash{{\SetFigFont{12}{14.4}{\familydefault}{\mddefault}{\updefault}{\color[rgb]{0,0,0}$\{\frac{l^\prime}{q}\alpha\}$}%
}}}}
\put(3046,-2510){\makebox(0,0)[lb]{\smash{{\SetFigFont{12}{14.4}{\familydefault}{\mddefault}{\updefault}{\color[rgb]{0,0,0}$\{\frac{l}{q}\alpha\}$}%
}}}}
\end{picture}%

}
\caption{Points with first component $\frac{l}{q}\alpha$}
\label{FIG:POINTSCIRCLE}
\end{center}
\end{figure}

We now consider two neighboring values of the first component, $\left\{\frac{l_1}{q}(\alpha)\right\}$ and $\left\{\frac{l_2}{q}(\alpha)\right\}$. Suppose $(m_1, n_1) \in M(t)$ has the largest value of $n$ among all points on the line $mq+np=l_1$ and $(m_2, n_2)~\in~M(t)$ has the smallest value of $n$ among all points on the line $mq+np=l_2$. Then, these two points form the spacing
\begin{eqnarray}
\frac{l_2-l_1}{q}\alpha - (n_1-n_2)\alpha\epsilon = \Delta_j(t) - (n_1-n_2)\alpha\epsilon.
\end{eqnarray}
Notice, however, that the line $mq+np=l_2$ goes through the x-axis in $R(t)$. Therefore, $n_2 \leq q-1$ or else $(m_2+p, n_2-q)$ satisfies $mq+np=l_2$ with a smaller $n$. We now consider $l_1 = kqp$, for all $1 \leq k \leq \frac{t}{q}$. Then, $(m_1,n_1) = (0,kq)$ and $n_2 \leq q-1$, so that each value of $a = n_1-n_2$ is unique as we vary $k$. Note also that $0 \leq a \leq t$.

We now show that if $a_1 \neq a_2$ then $\Delta_{j_1}(t)-a_1\epsilon\alpha \neq \Delta_{j_2}(t)-a_2\epsilon\alpha$. If $j_1 = j_2$, then it is clear. Otherwise, without loss of generality, suppose that $j_1 > j_2$, and for the sake of contradiction that $\Delta_{j_1}(t)-a_1\epsilon\alpha = \Delta_{j_2}(t)-a_2\epsilon\alpha$. Then note that $(a_1-a_2)\epsilon\alpha = \Delta_{j_1}(t)-\Delta_{j_2}(t)$. It is clear that $\Delta_{j_1}(t)~-~\Delta_{j_2}(t)$ can be expressed as $||l(\frac{\alpha}{q})||$ for some $-2stq \leq l \leq 2stq$. Therefore, 
\begin{eqnarray*}
\Delta_{j_1}(t)-\Delta_{j_2}(t)~\geq~\frac{K}{q(2stq)^\gamma}.
\end{eqnarray*}
 Furthermore, $t \geq a_1-a_2 > 0$. Thus, $t\epsilon\alpha \geq \frac{K}{q(2stq)^\gamma}$, giving $t \geq t_0$, a contradiction.

Thus, each $\Delta_j(t) - (n_1-n_2)\alpha\epsilon$ is unique, and since $1 \leq k \leq \frac{t}{q}$, this gives at least $\lfloor\frac{t}{q}\rfloor$ distinct spacings.
\end{proof}

We first use Lemma~\ref{LEM:INFSPACING} to construct vectors for which $D(t)$ grows fast:

\begin{prop}\label{PROP:FASTINFSPACING}
For any $0 < \delta \leq 1$, let $\alpha$ be Diophantine with exponent $\gamma$ and let
\begin{eqnarray}
\beta =\alpha\cdot\left(1+\displaystyle\sum_{n=0}^{\infty}\lambda^{-k^n}\right) \nonumber
\end{eqnarray}
for some integer $k > \frac{(1+\gamma)(1+\delta)}{\delta}$ and some integer $\lambda > 1$. Then, for $\vect{\omega}=(\alpha,\beta)$ and $R$ as in Figure~\ref{FIG:RTRIANGLE} with $s \geq \beta/\alpha$, we have that $\displaystyle\limsup_{t\rightarrow\infty}\left(\frac{D(t)}{t^{1-\delta}}\right) = \infty$.
\end{prop}
\begin{proof}
We first let $\epsilon_i = \displaystyle\sum_{n=i+1}^{\infty} \lambda^{-k^n}$. Note that
\begin{eqnarray}\label{EQ:FISEPSILON}
\epsilon_i \leq \lambda^{-k^{i+1}}\sum_{n=0}^{\infty}(\lambda^{-n}) = \frac{\lambda^{-k^{i+1}}}{1-\frac{1}{\lambda}}.
\end{eqnarray}
Then let $\frac{p_i}{q_i} = \frac{\beta}{\alpha}-\epsilon_i \leq s$. It can be seen that $q_i = \lambda^{k^i}$. Finally, let 
\begin{eqnarray}
t_{i} = \displaystyle\left(\frac{K}{q_i^{1+\gamma}(2s)^{\gamma}\alpha\epsilon_i}\right)^{\frac{1}{1+\gamma}}/2. \nonumber
\end{eqnarray}
Thus, we can apply Lemma~\ref{LEM:INFSPACING} to see that $\displaystyle\frac{D(t_i)}{t_i^{1-\delta}} \geq \frac{\lfloor t_i/q_i \rfloor}{t_i^{1-\delta}}$. Therefore, it suffices to show that $\lim(t_i) = \infty$ and that $\lim\left(\frac{t_i^\delta}{q_i}\right)=\infty$.

In order to show that $\lim(t_i) = \infty$, it suffices to prove that
$\lim(q_i^{1+\gamma}\epsilon_i)~=~0$. However, by (\ref{EQ:FISEPSILON}),
\begin{eqnarray}
q_i^{1+\gamma}\epsilon_i \leq
\frac{\lambda^{(1+\gamma)k^{i}-k^{i+1}}}{1-\frac{1}{\lambda}}, \nonumber
\end{eqnarray}
 which approaches $0$ since 
$k > \frac{(1+\gamma)(1+\delta)}{\delta} \geq 1+\gamma$.

Similarly, we show that $\lim\left(\frac{t_i^\delta}{q_i}\right)=\infty$ by
proving that $\lim(q_i^{(1+\gamma)(1+\delta)}\epsilon_i^\delta)~=~0$. Again, by
(\ref{EQ:FISEPSILON}), 
\begin{eqnarray}
q_i^{(1+\gamma)(1+\delta)}\epsilon_i^\delta \leq \lambda^{(1+\gamma)(1+\delta)k^{i}-(\delta)k^{i+1}}\left(\frac{1}{1-\frac{1}{\lambda}}\right)^\delta, \nonumber
\end{eqnarray}
which goes to $0$ since $k > \frac{(1+\gamma)(1+\delta)}{\delta}$.

\end{proof}
\begin{cor}\label{COR:INFSPACING}
There exist vectors $\vect{\omega}=(\alpha,\beta)$ that are Diophantine with exponent $5$ for which $\displaystyle\limsup_{t\rightarrow\infty}\left(D(t)\right) = \infty$ is satisfied for the region $R$ as shown in Figure~\ref{FIG:RTRIANGLE} (with $s \geq \beta/\alpha$).
\end{cor}
\begin{proof}
Let $\tau = \left(1+\displaystyle\sum_{n=0}^{\infty}\lambda^{-5^n}\right)$ for some integer $\lambda > 1$. Note that $\tau$ is Diophantine with exponent~$5$. Thus, by Corollary~\ref{COR:FULLMEASURE}, there exists a full measure set $\Lambda \subseteq [1,\infty)$ such that any $\alpha \in \Lambda$ makes $(\alpha, \alpha\tau)$ Diophantine with exponent $5$. Since the set $\Omega \in [1,\infty)$ of numbers Diophantine with exponent $5/4$ is also of full measure, the set $\Lambda \cap \Omega$ is full measure. Choosing $\alpha$ from $\Lambda \cap \Omega$, $\beta = \alpha\tau$, and $\delta = 1$ allows us to apply Proposition~\ref{PROP:FASTINFSPACING}, since $k = 5 > \frac{(1+\gamma)(1+\delta)}{\delta}$, which finishes the proof.
\end{proof}

Although we cannot answer Question~\ref{QUEST:NONBADLYAPPROXIMABLE} at the present time, one can easily use the Transference Theorems of \cite{CASSELS} to show:

\begin{prop}\label{PROP:NONBADLYAPPROXIMABLE}
If $\vect{\omega}=(\omega_1, \ldots, \omega_d)$ is not badly approximable, then there is a sequence $t_n\rightarrow\infty$ for which the ratio $\frac{\Delta_{D(t_n)}(t_n)}{\Delta_1(t_n)}$ goes to infinity.
\end{prop}

The following combination of Theorems 2 and 7 of \cite{CASSELS}, in the case when $m=d$ and $n=1$, will be used in the proof of  Proposition~\ref{PROP:NONBADLYAPPROXIMABLE}.

\begin{thm}\label{THM:CASSELSCOMBO}
For all $\gamma$ and $X$, let $\tilde{X}=\frac{\gamma^{-1/d}X}{4d^2}$ and $\tilde{\gamma}=\frac{X^{-d}\gamma^\frac{d-1}{d}}{4d}$.

\noindent If there exists $\vect{x}=(x_1,\ldots,x_d)\neq 0$ such that
        \[\left|\vect{x}\right|\leq X\ and\ \left\|\vect{x}\cdot\vect{\omega}\right\|\leq C=\gamma X^{-d},
\]
then there exists $\vect{\alpha}$, such that for all $\vect{x}$
        \[\left|\vect{x}\right|\leq \tilde{X} \Rightarrow \left\|\vect{x}\cdot\vect{\omega}-\alpha\right\|\geq\tilde{\gamma}.
\]
\end{thm}

\noindent
{\it Proof of Proposition~\ref{PROP:NONBADLYAPPROXIMABLE}: }
In this proof, we assume that the origin lies in the region, $R$, however, methods similar to those in the proof of Lemma~\ref{LEM:LARGESTSPACING} can be used to show the proposition is true when the origin does not lie in the region.

Similar to the proof of Theorem~\ref{THM:TRANSFERENCE}, let $A$ and $B$ be closed balls with radii $\rmin$ and $\rmax$ respectively, taken with respect to the infinity norm $|\cdot|_\infty$, such that $A \subseteq R \subseteq B$. Then, let $A(t) = \{xt \mid x \in A \}$ and $B(t) = \{xt \mid x \in B\}$.

Let $\gamma_n$ be a sequence going to 0.  Since $\vect{\omega}$ is not badly approximable, for each $\gamma_n$, there exist $\vect{x}_n$ such that $\left\|\vect{x}_n\cdot\vect{\omega}\right\|\leq\gamma_n\left|\vect{x}_n\right|_\infty^{-d}$.  Let $X_n:=\left|\vect{x}_n\right|_\infty$.  Then, by Theorem~\ref{THM:CASSELSCOMBO} above, there exists $\alpha_n$ such that for all $\vect{x}$
        \[\left|\vect{x}\right|_\infty\leq\frac{\gamma_n^{-1/d}X_n}{4d^2}\Rightarrow \left\|\vect{x}\cdot\vect{\omega}-\alpha_n\right\|\geq\frac{X_n^{-d}\gamma_n^\frac{d-1}{d}}{4d}.
\]

Let $t_n$ be the smallest $t$ such that $\vect{x}_n\subseteq\R{t_n}$. This also implies $\rmin t_n\leq\left|\vect{x}_n\right|_\infty$. Moreover, the smallest spacing,
        \[\Delta_{1}(t_n)\leq\left\|\vect{x}_n\cdot\vect{\omega}\right\|\leq\gamma_n\left|\vect{x}_n\right|_\infty^{-d}.
\]
Thus, for $n$ sufficiently large that $\frac{\rmax}{\rmin}\leq\frac{\gamma_n^{-1/d}}{4d^2}$, we have $\rmax t_n\leq\frac{\gamma_n^{-1/d}X_n}{4d^2}$.  This implies that, the largest spacing, $\Delta_{(t_n)}(t_n)$ satisfies 
\begin{eqnarray}
\frac{X_n^{-d}\gamma_n^\frac{d-1}{d}}{4d}\leq\Delta_{D(t_n)}(t_n).\nonumber
\end{eqnarray}
Therefore,
        \[\frac{\frac{X_n^{-d}\gamma_n^\frac{d-1}{d}}{4d}}{\gamma_n X_n^{-d}}=\frac{\gamma_n^{-1/d}}{4d}\leq\frac{\Delta_{D(t_n)}(t_n)}{\Delta_{1}(t_n)}.
\]
  Since $\gamma_n\rightarrow 0$, this implies $\frac{\Delta_{D(t_n)}(t_n)}{\Delta_{1}(t_n)}\rightarrow\infty$.


\appendix
\normalsize
\section{Selecting Diophantine Vectors}
\begin{prop}\label{PROP:FULLMEASURE}
Let $\tau$ be Diophantine with exponent $\kappa$.  There exists a full measure $\Lambda\subseteq[0,1]$ such that if $\alpha\in\Lambda$, then $(\tau,\alpha)$ is Diophantine with exponent $\gamma$, for any $\gamma>\max(3,\kappa)$.
\end{prop}

\begin{proof}
Notice that $(\tau,\alpha)$ is Diophantine with exponent $\gamma$ if and only it $\exists K$ such that
\begin{eqnarray}
|m+n\tau+p\alpha|\geq\frac{K_0}{|(m,n,p)|^\gamma} \nonumber
\end{eqnarray}
for all nonzero integer vectors $(m,n,p)$.

If $p=0$, then $|m+n\tau+p\alpha|=|m+n\tau|\geq\frac{K_0}{|(m,n)|^\kappa}\geq\frac{K_0}{|(m,n,p)|^\gamma}$, for all $\alpha$, since $\tau$ is Diophantine with exponent $\kappa$.

If $p\neq0$, the plane in $\mathbb{R}^3$ orthogonal to $(m,n,p)$ at the origin intersects the line
        \[L:\left\{x=1, y=\tau\right\}
\]
at angle $\phi$, where $\sin\phi=\frac{p}{|(m,n,p)|}$.

Denote $L_0$ as the portion of $L$, where $0\leq z\leq1$.

We will now overestimate the length of $L_0$ that is hit by a given slab,
        \[S_{(m,n,p)}:=\left\{(x,y,z)\,\vline\,|mx+ny+pz|\leq\frac{K}{|(m,n,p)|^{\gamma}}\right\}.
\]
Let $P_{(m,n,p)}$ be the plane orthogonal to $(m,n,p)$ .  Since $P_{(m,n,p)}$ and $L_0$ intersect at angle $\phi$, with $\sin\phi=\frac{p}{|(m,n,p)|}$, the distance in $L$ between the upper plane of the slab $S_{(m,n,p)}$ and the plane $P_{(m,n,p)}$ is
        \[d_{(m,n,p)}:=\frac{K}{|(m,n,p)|^{\gamma+1}}\cdot\frac{1}{|\sin\phi|};
\]
\noindent
See Figure~\ref{FIG:SLAB}.
\begin{figure}[hbtp]
\begin{center}
\scalebox{0.8}{

\begin{picture}(0,0)%
\includegraphics{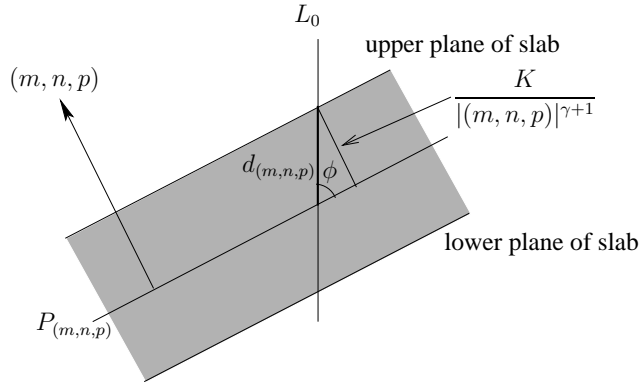}%
\end{picture}%
%
%
\setlength{\unitlength}{4144sp}%
\begingroup\makeatletter\ifx\SetFigFont\undefined%
\gdef\SetFigFont#1#2#3#4#5{%
  \reset@font\fontsize{#1}{#2pt}%
  \fontfamily{#3}\fontseries{#4}\fontshape{#5}%
  \selectfont}%
\fi\endgroup%
\begin{picture}(4931,2892)(1501,-4480)
\put(3646,-1771){\makebox(0,0)[lb]{\smash{{\SetFigFont{12}{14.4}{\rmdefault}{\mddefault}{\updefault}{\color[rgb]{0,0,0}$L_0$}%
}}}}
\put(3860,-2938){\makebox(0,0)[lb]{\smash{{\SetFigFont{12}{14.4}{\rmdefault}{\mddefault}{\updefault}{\color[rgb]{0,0,0}$\phi$}%
}}}}
\put(1516,-2251){\makebox(0,0)[lb]{\smash{{\SetFigFont{12}{14.4}{\rmdefault}{\mddefault}{\updefault}{\color[rgb]{0,0,0}$(m,n,p)$}%
}}}}
\put(1716,-4081){\makebox(0,0)[lb]{\smash{{\SetFigFont{12}{14.4}{\rmdefault}{\mddefault}{\updefault}{\color[rgb]{0,0,0}$P_{(m,n,p)}$}%
}}}}
\put(4835,-2388){\makebox(0,0)[lb]{\smash{{\SetFigFont{12}{14.4}{\rmdefault}{\mddefault}{\updefault}{\color[rgb]{0,0,0}$\displaystyle\frac{K}{|(m,n,p)|^{\gamma+1}}$}%
}}}}
\put(3257,-2898){\makebox(0,0)[lb]{\smash{{\SetFigFont{12}{14.4}{\rmdefault}{\mddefault}{\updefault}{\color[rgb]{0,0,0}$d_{(m,n,p)}$}%
}}}}
\end{picture}%

}

\caption{The intersection of the line segment $L_0$ and the slab $S_{(m,n,p)}$.}
\label{FIG:SLAB}
\end{center}
\end{figure}

Thus, the total length of $L$ that intersects $S_{(m,n,p)}$ is less than or equal to
\begin{eqnarray}  \frac{2K}{|(m,n,p)|^{\gamma+1}}\cdot\frac{1}{|\sin\phi|}=\frac{2K}{|(m,n,p)|^{\gamma+1}}\cdot\frac{|(m,n,p)|}{|p|}=\frac{2K}{|(m,n,p)|^{\gamma}|p|}\leq\frac{2K}{|(m,n,p)|^{\gamma}}. \nonumber
\end{eqnarray}
Since $\gamma>3$, $\displaystyle\sum_{(m,n,p)\in\mathbb{Z}^3\sm\left\{0\right\}}\frac{2K}{|(m,n,p)|^{\gamma}}$ is a converging series.  Thus,
        \[\textrm{length}\left(\bigcup_{(m,n,p)\in\mathbb{Z}^3\sm\left\{0\right\}}(S_{(m,n,p)}\cap L)\right)\leq JK,
\]
where $J$ is some constant.  Therefore,

\begin{eqnarray}
        \Omega=\bigcap_{K>0}\bigcup_{(m,n,p)\in\mathbb{Z}^3\sm\left\{0\right\}}S_{(m,n,p)}\cap L_0 \nonumber
\end{eqnarray}
has measure zero.  Since $\Lambda$ is the complement to $\Omega$, $\Lambda$ is a full measure subset of $[0,1]$.
\end{proof}

\begin{cor}\label{COR:FULLMEASURE}
If $\tau$ is Diophantine with exponent $\kappa$, then there is a full measure set $\Lambda\subseteq[1,\infty]$ such that if $\alpha\in\Lambda$, then $(1,\alpha,\alpha\tau)$ is Diophantine with exponent $\gamma$, for any $\gamma\geq\max{3,\kappa}$.
\end{cor}
\begin{proof}
This follows by dividing $|m+n\tau+p\alpha|\geq\frac{K}{|(m,n,p)|^\gamma}$ by $\alpha$.
\end{proof}

\vspace{0.2in}

\textbf{Acknowledgements.}
This paper is the culmination of the work of Homma, Ji, and Shen for the 2010
Siemens Competition. The project was designed and mentored by Bleher and
Roeder. We thank the Siemens Foundation for their kind hospitality. We thank
Freeman Dyson for providing Bleher with the preprint \cite{DYSON} together with
the letters from Michael Boshernitzan \cite{BOSHERNITZAN1,BOSHERNITZAN2}.   
In addition, Rodrigo P\'{e}rez has provided many valuable comments.   Finally, we thank the referee for his or her helpful remarks.

The work of Bleher is
supported in part by the National Science Foundation grants DMS-0652005 and DMS-0969254. The work of
Roeder was partially supported by startup funds from the Department of
Mathematics at IUPUI.



\end{document}